\documentclass[preprint,amsmath,amssymb]{revtex4}
\usepackage{epsfig}

\begin{document}

\def\unit#1{{\rm #1}}

\title{Testing the Newton Law at Long Distances}
\author{Serge Reynaud}
\affiliation{Laboratoire Kastler Brossel,
UPMC case 74, Jussieu, F75252 Paris Cedex 05}
\thanks{Laboratoire de l'Ecole
Normale Sup\'erieure, de l'Universit\'e Pierre et Marie Curie et du CNRS}
\author{Marc-Thierry Jaekel}
\affiliation{Laboratoire de Physique Th\'eorique de l'ENS,
24 rue Lhomond, F75231 Paris Cedex 05}
\thanks{Laboratoire du CNRS, de l'Ecole Normale
Sup\'erieure et de l'Universit\'e Paris-Sud}
\date{9 Nov 2004}

\begin{abstract}
Experimental tests of Newton law put stringent constraints on potential
deviations from standard theory with ranges from the millimeter to the size
of planetary orbits.\ Windows however remain open for short range
deviations, below the millimeter, as well as long range ones, of the order
of or larger than the size of the solar system. We discuss here the relation
between long range tests of the Newton law and the anomaly recorded on
Pioneer 10/11 probes.
\end{abstract}
\maketitle

\section{Introduction}

The validity of Newton force law has been tested at distances ranging from
the millimeter in laboratory experiments\cite{Adelberger03} to the size of
planetary orbits.\cite{Fischbach98} But windows remain open for violations
of the inverse square law at short ranges, below the millimeter, as well as
long ones, of the order of or larger than the size of the solar system. This
is also true for tests of general relativity which tightly constrain
potential violations in the solar system\cite{Will01} but let open space
for deviations at small or large scales.

The accuracy of short range tests has recently shown impressive progress,
for gravity experiments pushed to smaller distances\cite%
{Hoyle,Long,Chiaverini} as well as for Casimir force experiments.\cite%
{Lamoreaux,Mohideen,Ederth,Chan,Bressi,Decca} In both cases, the agreement
between theory and experiment is good. For Casimir experiments, it reaches
an accuracy near the 1\% level, after having accounted for the effects of
imperfect reflection of the metallic mirrors used in the experiments.\cite%
{Bordag01,Lambrecht02} The agreement can then be translated into
constraints on potential violations of Newton force law at ranges from
nanometer to millimeter.\cite{Decca03,Chen04}

On the other hand, long range tests of the Newton law are performed by
monitoring the motions of planets or probes in the solar system. The tests
bearing on the third Kepler law or the precession of the perihelion of
planets\cite{Talmadge88} confirm the validity of general relativity.\cite%
{Coy03} The accuracy is especially good for ranges of the order of the
Earth-Moon\cite{Williams96} or Sun-Mars distances 
(see for example Refs.~20, 21, 22, 23). However, the Doppler
data recorded on the Pioneer 10/11 probes show an anomaly when compared with
calculations based on general relativity.\cite{Anderson98} The anomaly may
be represented as an anomalous acceleration directed towards the Sun with a
roughly constant amplitude.\cite{Anderson02} It has not been explained to
date though a number of mechanisms have been considered to this aim (see the
discussions and references in Refs.~26, 27, 28, 29).

At even larger scales, the rotation curves of galaxies show a conflict with
general relativity as soon as the source of gravity is identified with the
matter detected by electromagnetic means. Due to the excellent agreement of
gravity tests with general relativity, \ this anomaly is commonly accounted
for by introducing \textquotedblleft dark matter\textquotedblright\
components designed to fit the rotation curves while keeping the gravity
laws untouched. As long as dark matter components are not detected by other
means, the anomaly can as well be ascribed to modifications of gravity at
galactic scales.\cite{Goldman04,Mannheim97,Milgrom02,Sanders02,Aguirre01ff}
Similar statements apply for the observation of an accelerated expansion
through the relation between redshifts and luminosities for supernovae. This
observation can be given alternative descriptions in terms either of
\textquotedblleft dark energy\textquotedblright\ or of modified gravity at
cosmic scales.\cite{Carroll04} Note that modifications of gravity are
expected to be produced by vacuum induced effects\cite{Sakharov} or by
effective gravity at low energy deduced from unification models.\cite%
{Dvali03,Reuter04} An important requirement to be met by any such
modification of gravity is that it is compatible with observations on
galactic or cosmic scales while still matching the strict bounds set by
gravity tests in the solar system.

The Pioneer anomaly may be a central piece of information in this context by
pointing at some anomalous behaviour of gravity at scales of the order of
the size of the solar system. In the following, we focus our attention on
the key question of compatibility of the Pioneer anomaly with other gravity
tests. After having briefly recalled the observations, we consider the idea
that the anomaly could be explained simply from a long-range deviation from
the Newton potential, for example with a Yukawa form. We show that this
explanation cannot be upheld against the data known for planetary tests of
Newton law. More precisely, if the anomalous acceleration is ascribed to a
Yukawa deviation from Newton law, the deviation is so large that it cannot
remain unnoticed on the motions of outer planets, primarily Mars. This
conclusion was already drawn\cite{AndersonYukawa} but it is written here
under a form allowing us to put emphasis on the challenge raised by the
incompatibility. The present paper also prepares discussions of a
modification of Einstein theory of gravity to be presented 
elsewhere.\cite{JaekelTBP}

\section{Pioneer Anomaly}

The anomaly is recorded on radio tracking data from the Pioneer 10/11 probes
during their travel to the outer parts of the solar system. At distances $r$
from the Sun between 20 and 70 astronomical units ($\mathrm{AU}$), the
Doppler data have shown a deviation from calculations based on general
relativity. The anomaly is observed as a linear variation with time of the
Doppler residuals, that is the differences of the observed Doppler velocity
from the modelled Doppler velocity (see Fig.~8 of Ref.~25). It may
be represented as an anomalous acceleration directed towards the Sun with a
roughly constant amplitude on the range of distances over which it has been
detected 
\begin{equation}
a_{P}\simeq 8\times 10^{-10}\unit{m}\unit{s}^{-2}\qquad ,\qquad 20~\mathrm{AU%
}<r<70~\mathrm{AU}  \label{Pioneer}.
\end{equation}%
The anomaly can also be represented as a clock acceleration with the
striking feature that its value $a_{P}/c\simeq 3\times 10^{-18}\unit{s}\unit{%
s}^{-2}$ is nearly equal to the Hubble frequency with a value $H\sim 80\unit{%
km}\unit{s}^{-1}\mathrm{Mpc}^{-1}$ for the Hubble constant.\cite{Anderson02}

Though a number of mechanisms have been considered to this aim,\cite%
{Anderson02b,Anderson03,Nieto04,Turyshev04} no satisfactory explanation of
the anomalous signal has been found to date. Potential systematic effects do
not seem to be able to reach the magnitude of the observed anomaly. Present
knowledge of the outer part of the solar system does apparently preclude
interpretations in terms of gravity or drag effects of ordinary matter. The
inability of explaining the anomaly with conventional physics has given rise
to a growing number of new theoretical explanations. It has also motivated
proposals for new missions designed to study the anomaly and try to
understand its origin (see the references in Refs.~26, 27, 28, 29).

The importance of the Pioneer anomaly for fundamental physics and space
navigation certainly justifies it to be submitted to further scrutiny. On
the theoretical side, the incompatibility of the Pioneer anomaly with other
gravity tests appears to be a key question. We now discuss this question by
considering the possibility that the anomaly could be explained from a
long-range deviation from the Newton potential. To this aim, we use the
common model of a Newton potential modified through the addition of a Yukawa
perturbation.\cite{Fischbach98}

\section{Yukawa Modification of Newton Law}

\label{sec_Yukawa}Throughout the paper, the potential energy $V$ is written
as the product of the mass energy of the probe by a dimensionless potential 
$V\left( r\right) \equiv mc^{2}\Phi \left( r\right) $.
The latter is the sum of the Newton potential and of a Yukawa correction 
\begin{equation}
\Phi \left( r\right) =-\frac{G_{\infty }M}{rc^{2}}\left( 1+\alpha _{\infty
}e^{-\frac{r}{\lambda }}\right)   \label{Yukaw},
\end{equation}%
$G_{\infty }$ is the effective Newton constant at large distances and $M$
the mass of the Sun; $\lambda $ is the range of the Yukawa potential and $%
\alpha _{\infty }$ its amplitude measured with respect to $G_{\infty }$. The
influence of the Yukawa perturbation disappears at the long distance limit $%
r\gg \lambda $ \ but it is significant otherwise. In particular, it gives
rise to a potential linear in the distance $r$ in the domain $r\ll \lambda $
and, then, to a constant anomalous acceleration (see the next section). In
the general case, the acceleration may be written as 
\begin{equation}
A\left( r\right) \equiv -c^{2}\frac{\partial \Phi }{\partial r}=-\frac{%
G_{\infty }M}{r^{2}}\left[ 1+\alpha _{\infty }e^{-\frac{r}{\lambda }}\left(
1+\frac{r}{\lambda }\right) \right]   \label{Accel}.
\end{equation}%
The anomalous acceleration contained in this expression could account for
the Pioneer anomaly, if the range $\lambda $ is larger than $70~\mathrm{AU}$%
. But the value of the correction would thus be too large to remain
unnoticed on planetary tests. In the following, we will put this
incompatibility under a more precise form.

Before going along this discussion, we rewrite the modified potential (\ref%
{Yukaw}) in terms of a running gravitational constant. To this aim, we write
the Laplacian of its two components which obey respectively a Poisson
equation (massless field propagation) and a Yukawa equation (massive field
propagation) 
\begin{equation}
\Delta \left( -\frac{1}{r}\right) =4\pi \delta (\mathbf{x})\qquad ,\qquad
\left( \Delta -\frac{1}{\lambda ^{2}}\right) \left( -\frac{e^{-\frac{r}{%
\lambda }}}{r}\right) =4\pi \delta (\mathbf{x}).
\end{equation}%
The right hand sides represent point sources with $\delta (\mathbf{x})$ the
Dirac distribution\ for 3-dimensional space position $\mathbf{x}$. This
equation leads to an expression of $\Phi $ in Fourier space as a function of
the spatial part $\mathbf{k}$ of the wavevector 
\begin{equation}
-\mathbf{k}^{2}\Phi \left[ \mathbf{k}\right] =4\pi \frac{G_{\infty }M}{c^{2}}%
\left( 1+\alpha _{\infty }\frac{\mathbf{k}^{2}}{\mathbf{k}^{2}+\frac{1}{%
\lambda ^{2}}}\right). 
\end{equation}%
This expression may equivalently be written in terms of a running constant
which replaces the Newton constant in the Poisson law 
\begin{equation}
-\mathbf{k}^{2}\Phi \left[ \mathbf{k}\right] \equiv 4\pi \frac{\widetilde{G}%
\left[ \mathbf{k}\right] M}{c^{2}}\qquad ,\qquad \widetilde{G}\left[ \mathbf{%
k}\right] =G_{\infty }\left( 1+\alpha _{\infty }\frac{\mathbf{k}^{2}}{%
\mathbf{k}^{2}+\frac{1}{\lambda ^{2}}}\right)   \label{Running}.
\end{equation}%
The running constant $\widetilde{G}\left[ \mathbf{k}\right] $ has been
defined as the function of momentum and it goes from the value $G_{\infty }$
for small values of $\mathbf{k}\lambda $ to the value $G_{\infty }\left(
1+\alpha _{\infty }\right) $ for large values of $\mathbf{k}\lambda $. Note
that $\widetilde{G}\left( r\right) $, which can be defined from $\widetilde{G%
}\left[ \mathbf{k}\right] $ through an inverse Fourier transform, obeys $%
\Delta \Phi \left( r\right) =4\pi \frac{\widetilde{G}\left( r\right) M}{c^{2}%
}$ (but not $\Phi \left( r\right) =-\frac{\widetilde{G}\left( r\right) M}{%
rc^{2}}$).

\section{Long Range Limit}

\label{sec_LongRange}In the following, we will consider with a particular
attention the long range limit, that is the domain $r\ll \lambda $ which
could provide us with an explanation of the nearly constant value of the
anomalous acceleration. To this purpose, we introduce an expansion of the
modified potential (\ref{Yukaw}) better adapted to this domain 
\begin{equation}
\Phi \left( r\right) =-\frac{GM}{rc^{2}}+\delta \Phi \left( r\right) \qquad
,\qquad \delta \Phi \left( r\right) =\frac{GM\alpha }{rc^{2}}\left( 1-e^{-%
\frac{r}{\lambda }}\right)   \label{YukawAnom}.
\end{equation}%
The gravitational constant and Yukawa amplitude have been redefined 
\begin{equation}
G\equiv G_{\infty }\left( 1+\alpha _{\infty }\right) \qquad ,\qquad G\alpha
\equiv G_{\infty }\alpha _{\infty }\qquad ,\qquad \alpha =\frac{\alpha
_{\infty }}{1+\alpha _{\infty }}  \label{redefG}.
\end{equation}%
Note that $G$ is the effective gravitational constant in the domain $r\ll
\lambda $. A power expansion of (\ref{YukawAnom}) in terms of $r/\lambda $
leads to 
\begin{equation}
\delta \Phi \left( r\right) \simeq \frac{GM\alpha }{c^{2}\lambda }\left[ 1-%
\frac{r}{2\lambda }+O\left( \frac{r^{2}}{\lambda ^{2}}\right) \right] 
\label{YukawExpans}.
\end{equation}%
The first term is constant and has no effect while the second one produces a
constant anomalous acceleration.

In general, the anomalous acceleration can be defined from (\ref{Accel}) as\ 
\begin{equation}
A\left( r\right) =-\frac{GM}{r^{2}}+\delta A\left( r\right) \qquad ,\qquad
\delta A\left( r\right) =\frac{GM\alpha }{r^{2}}\left[ 1-e^{-\frac{r}{%
\lambda }}\left( 1+\frac{r}{\lambda }\right) \right]   \label{AccelAnom}.
\end{equation}%
It is then expanded in the domain $r\ll \lambda $ as 
\begin{equation}
\delta A\left( r\right) \simeq \frac{GM\alpha }{\lambda ^{2}}\left[ \frac{1}{%
2}-\frac{r}{3\lambda }+O\left( \frac{r^{2}}{\lambda ^{2}}\right) \right] 
\label{AccelExpans}.
\end{equation}%
As already stated, the anomalous acceleration $\delta A$ is essentially a
constant in this domain. \ Should we identify it with the anomalous
acceleration observed on the Pioneer probes, we would obtain an
interpretation of the Pioneer anomaly. The sign convention is
such that $\delta A<0$ is needed to fit the observed anomaly. Note that the
constant value of $\delta A$ depends only on the combination $\alpha
/\lambda ^{2}$ of the two parameters entering the Yukawa potential. Hence,
there is a degeneracy in the choice of these two parameters which can be
characterized by the relation 
\begin{equation}
\frac{\alpha }{\lambda ^{2}}=-\frac{1}{\Lambda ^{2}}\qquad ,\qquad \Lambda =%
\sqrt{\frac{GM}{2a_{P}}}\simeq 6300\mathrm{AU}\simeq 10^{15}\unit{m}
\label{degenAlphaLambda}.
\end{equation}%
$\Lambda $ is the distance from the Sun at which the Newton acceleration is
twice the Pioneer anomalous acceleration. When plotted on a $\left( \lambda
,\alpha \right) $ diagram with log-log coordinates, the relation (\ref%
{degenAlphaLambda}) defines a straight line with a slope 2. Note that only
points with $\lambda >70~\mathrm{AU}$ fit a constant anomalous acceleration
on the range of distances probed by Pioneer 10/11. Now the half-line thus
defined $\left( 70~\mathrm{AU}<\lambda ,\alpha =-\frac{\lambda ^{2}}{\Lambda
^{2}}\right) $ is excluded by existing planetary tests of Newton law.\cite%
{Fischbach98} The point is illustrated on Fig.~1 which shows recently
updated constraints drawn from the analysis of the motions of planets and
probes in the solar system:\cite{Coy03} the values $\left( \lambda ,\alpha
\right) $ fitting the anomaly are clearly in the forbidden domain. This
conflict will be written directly in terms of the anomalous acceleration $%
\delta A$ in the next sections.

\begin{figure}
\centerline{\psfig{file=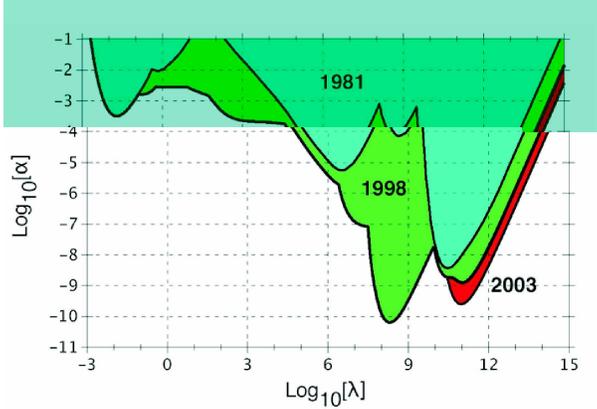,width=8cm}}
\vspace*{8pt}
\caption{Constraints on a Yukawa modification of Newton law.
The colored domains represent, on a log-log plot, the excluded values 
for amplitude $\alpha$ as a function of the range $\lambda$ (measured in m).
Courtesy of the authors of Ref.~18.}
\end{figure}

The discussion can also be translated into terms bearing on the running
constant. To this aim, we rewrite (\ref{Running}) under a form better
adapted to the domain $\mathbf{k}\lambda \gg 1$ (which corresponds to $r\ll
\lambda $) 
\begin{equation}
\widetilde{G}\left[ \mathbf{k}\right] =G+\delta \widetilde{G}\left[ \mathbf{k%
}\right] \qquad ,\qquad \delta \widetilde{G}\left[ \mathbf{k}\right] =-\frac{%
G\alpha }{\mathbf{k}^{2}\lambda ^{2}+1}  \label{RunningAnom}.
\end{equation}%
We then expand the anomalous part in this domain 
\begin{equation}
\delta \widetilde{G}\left[ \mathbf{k}\right] =-\frac{G\alpha }{\mathbf{k}%
^{2}\lambda ^{2}}\left[ 1+O\left( \frac{1}{\mathbf{k}^{2}\lambda ^{2}}%
\right) \right]   \label{RunningExpans}.
\end{equation}%
As the constant anomalous acceleration $\delta A$ in (\ref{AccelExpans})
with which it is directly associated, the anomalous term $\delta \widetilde{G%
}\left[ \mathbf{k}\right] $ is proportional to the combination $\alpha
/\lambda ^{2}$. Note that Eqs.~(\ref{YukawExpans}), (\ref%
{AccelExpans}) and (\ref{RunningExpans}), valid in the domain $r\ll \lambda $ (or
equivalently $\mathbf{k}\lambda \gg 1$), are sufficient for the purpose of
our discussions. At the same time, they have been derived from Eqs.~(%
\ref{Yukaw}), (\ref{Accel}) and (\ref{Running}) which are better behaved at the limit
of large distances $r\gg \lambda $ (or equivalently $\mathbf{k}\lambda \ll 1$%
).

\section{Planetary Tests}

\label{sec_planetary}Tests of the Newton law bearing either on the third
Kepler law or on the precession of the perihelion of planets are known to
confirm the validity of general relativity with a good accuracy. We now make
the significance of this statement more explicit by writing directly in
terms of the anomalous acceleration $\delta A$ the constraints drawn from
these tests. Since we study gravity in the outer solar system, we use the
Newton theory with the Sun described as a motionless point source.
The motion of the probe mass takes place in the
central potential (\ref{Yukaw}) and we can write the conservation of energy $%
E$ and angular momentum $J$ 
\begin{eqnarray}
\frac{E}{mc^{2}} &=&\frac{\upsilon _{r}^{2}+\upsilon _{\varphi }^{2}}{2c^{2}}%
+\Phi \left( r\right) \qquad ,\qquad \frac{J}{m}=r\upsilon _{\varphi } 
\notag \\
\upsilon _{r} &=&\frac{dr}{dt}\qquad ,\qquad \upsilon _{\varphi }=r\frac{%
d\varphi }{dt},
\end{eqnarray}%
$r$ is the distance from the Sun, $t$ the time coordinate and $\varphi $ the
azimutal angle; the trajectory has been assumed to stay in the plane $\theta
={\frac{\pi }{2}}$. When eliminating time, we obtain the equation of motion
as 
\begin{equation}
\frac{d^{2}u}{d\varphi ^{2}}+u=-\frac{m^{2}}{J^{2}}r^{2}A\left( r\right)
\qquad ,\qquad u\equiv \frac{1}{r}  \label{eqMot}.
\end{equation}%
Note that $r^{2}A\left( r\right) $ can be replaced by $c^{2}\frac{\partial
\Phi }{\partial u}$ when $\Phi $ is the potential expressed in terms of the
variable $u$.

Using the standard Newton law for the potential ($\alpha =0$), the right
hand side in (\ref{eqMot}) is a constant and the Kepler ellipse is recovered 
\begin{equation}
u=\frac{1+e\mathrm{\cos }\varphi }{p}\qquad ,\qquad p\equiv a\left(
1-e^{2}\right) =\frac{J^{2}}{GMm^{2}}\qquad ,\qquad E=-\frac{GMm}{2a}
\label{ellipse}.
\end{equation}%
The orbital frequency $\omega =\frac{2\pi }{T}$, with $T$ the orbital
period, is given by the third Kepler law $\omega ^{2}a^{3}=GM$ or,
equivalently, by an expression directly drawn from (\ref{eqMot}) 
\begin{equation}
\left[ u\right] _{\mathrm{st}}=-\frac{m^{2}}{J^{2}}\left[ r^{2}A\right] _{%
\mathrm{st}}=\frac{GMm^{2}}{J^{2}}  \label{KeplerSt}.
\end{equation}%
Here $\left[ u\right] _{\mathrm{st}}$ and $\left[ r^{2}A\right] _{\mathrm{st}%
}$ are the values obtained for $u$ and $r^{2}A$ in standard theory.

With the modified Newton law ($\alpha \neq 0$), there is a correction to the
third Kepler law (\ref{KeplerSt}). If we evaluate it on circular orbits\ ($%
e=0$), we obtain in a linear approximation with respect to the small
perturbation 
\begin{equation}
\frac{u}{\left[ u\right] _{\mathrm{st}}}=\frac{r^{2}A}{\left[ r^{2}A\right]
_{\mathrm{st}}}=1-\frac{r^{2}\delta A\left( r\right) }{GM}  \label{anomU}.
\end{equation}%
Consider now a planet, say Mars, for which the elements $a$ or $p$
describing the orbit have been measured through optical means, that is
independently of the measurement of the orbital period.\cite{Talmadge88}
The result obtained for Mars can then be compared with the reference value
fixed by the orbit of Earth by forming the ratio $u_{\mathrm{Mars}}/u_{%
\mathrm{Earth}}$ whose difference to unity now depends on the variation of (%
\ref{anomU}) from Earth to Mars. Defining the relative accuracy $\varepsilon
\equiv \left\vert \delta a/a\right\vert $ on the element $a$ of Mars as it
is measured in AU, we deduce that an anomalous acceleration could be noticed
under the condition 
\begin{equation}
\varepsilon <\left\vert \left[ \frac{a^{2}\delta A\left( a\right) }{GM}%
\right] _{\mathrm{Earth}}-\left[ \frac{a^{2}\delta A\left( a\right) }{GM}%
\right] _{\mathrm{Mars}}\right\vert   \label{condEps}.
\end{equation}%
Using (\ref{Accel}), this condition is read as a minimum value for the
amplitude 
\begin{equation}
\frac{\alpha }{\varepsilon }>\frac{1}{\left\vert \left[ \left( 1+\frac{a}{%
\lambda }\right) e^{-\frac{a}{\lambda }}\right] _{\mathrm{Mars}}-\left[
\left( 1+\frac{a}{\lambda }\right) e^{-\frac{a}{\lambda }}\right] _{\mathrm{%
Earth}}\right\vert }  \label{KeplerAccur}.
\end{equation}%
The curve showing the frontier of the domain is plotted on Fig.~2 with $\log
_{10}-\log _{10}$ scales. The range $\lambda $ as well as the radii are
measured in AU ($a=1$ for Earth and $1.5$ for Mars). The long range
asymptote on the curve corresponds to a fixed value for $\alpha /\lambda ^{2}
$ and, therefore, for the anomalous acceleration $\delta A$. This point will
be discussed in more detail below. 

\begin{figure}
\centerline{\psfig{file=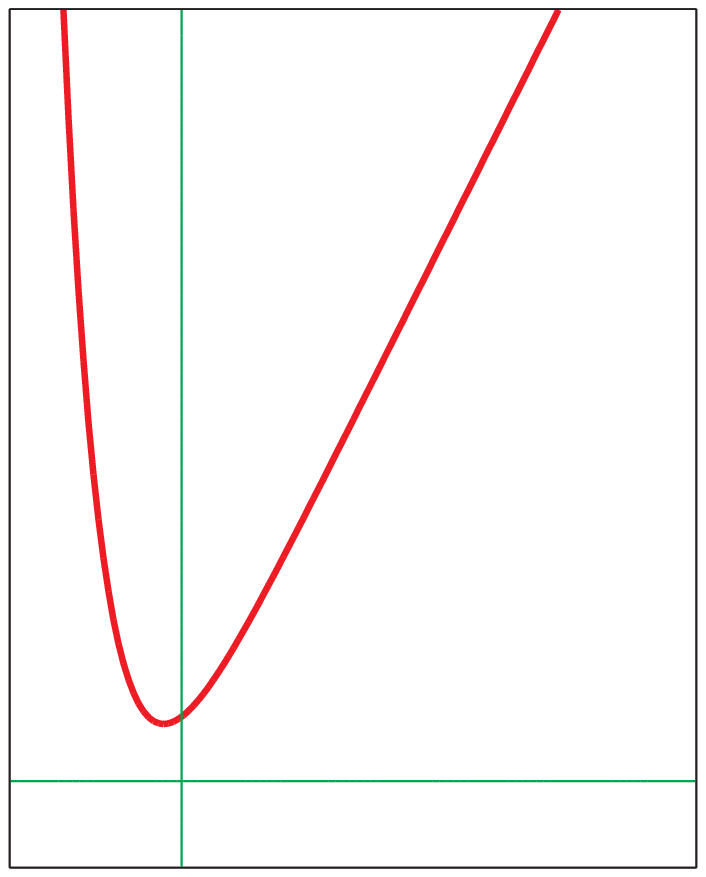,width=3.2cm} \hspace*{2cm}
\psfig{file=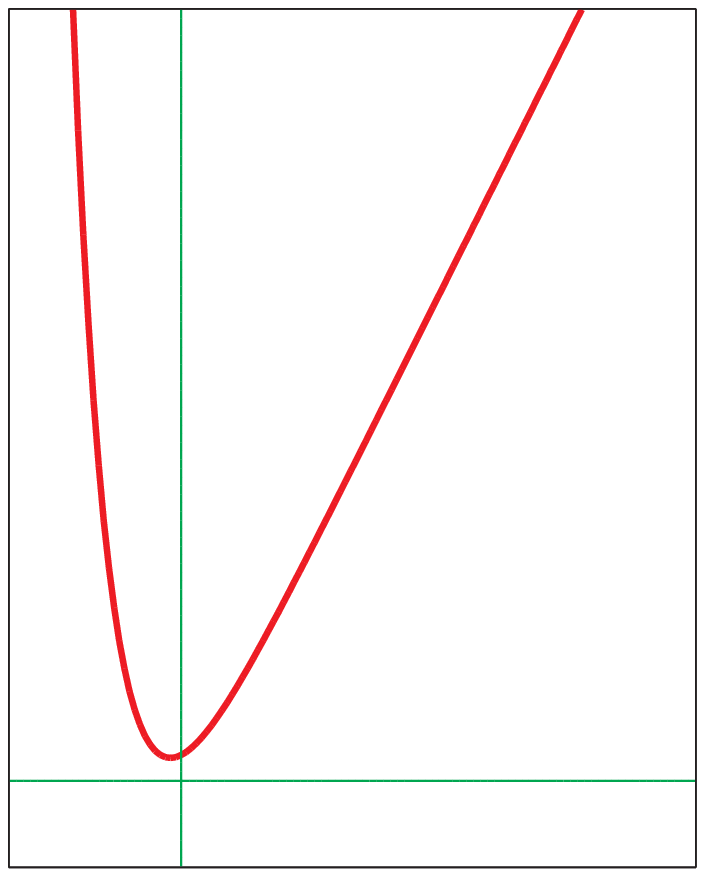,width=3.2cm} }
\vspace*{8pt}
\caption{Sensitivity of a Kepler test (left hand part) and precession
test (right hand part).
The full lines delineate the exclusion domain (above the curve)
for the value of the reduced amplitude $\alpha /\varepsilon$  
(definitions of $\varepsilon$  given in the text)
as a function of $\lambda$ (measured in AU).
The logarithms go from -2 to +6 on the horizontal scale 
and from -1 to +9 on the vertical scale with the zeros
indicated by thin lines.}
\end{figure}

Precise planetary tests are also performed by following the precession of
the perihelion of planets.\cite{Talmadge88,Fischbach98} For simplicity, we
evaluate it for an orbit with a low eccentricity $e\ll 1$ in a linear
approximation with respect to the small perturbation. In this case, the
variable $u$ undergoes a small sinusoidal variation around a constant $%
1/p\simeq 1/a$ (see Eq.~\ref{ellipse}). The equation of motion (\ref{eqMot})
can thus be replaced by an approximation corresponding to the linearization
of the associated variation of $r^{2}A\left( r\right) $ 
\begin{eqnarray}
&&\frac{d^{2}u}{d\varphi ^{2}}+\left( 1-2\delta\kappa \right) 
\left( u-\frac{1}{P}
\right) =0 \nonumber\\ 
&&\delta \kappa =\frac{1}{2}\frac{m^{2}}{J^{2}}\left[ r^{2}%
\frac{\mathrm{d}}{\mathrm{d}r}\left( r^{2}A\left( r\right) \right) \right]
_{r=a}=\frac{\alpha a^{2}}{2\lambda ^{2}}e^{-\frac{a}{\lambda }}.
\end{eqnarray}%
The parameter $1/p$ has been replaced by a modified value $1/P$ but this
does not matter for our present purpose. What is important for the
evaluation of the precession is the modification of the coefficient in front
of $u$. It implies that the perihelion (maximum value of $u$) is recovered
when $\varphi$ has ran over $2\pi \left(1+\delta \kappa \right)$ rather than the
standard value $2\pi $. Defining the relative accuracy 
$\varepsilon = \left\vert \delta \varphi / 2\pi \right\vert $ 
for a test of the precession,
we obtain the following relation for the anomaly to be detectable 
\begin{equation}
\varepsilon < \alpha \left[ \frac{a^{2}%
}{2\lambda ^{2}}e^{-\frac{a}{\lambda }}\right] _{\mathrm{Mars}}
\label{precessAccur}.
\end{equation}%
The frontier of the exclusion domain is drawn as the right hand plot on 
Fig.~2. 
It has roughly the same shape as the left hand plot and, in particular,
the long range asymptote on the curve corresponds again to a fixed value for 
$\alpha /\lambda ^{2}$. 

\section{Discussion}

\label{sec_Discussion}As already stated, the Yukawa correction of the Newton
law would produce a constant anomalous acceleration over the range of
distances probed by Pioneer 10/11 provided the Yukawa range is large enough $%
\lambda >70~\mathrm{AU}$. This means that the tests performed on planets or
probes in the solar system correspond to the limit of long ranges $r\ll
\lambda $. As shown in the preceding section, this also means that they test
a single combination $\alpha /\lambda ^{2}$ of the two parameters entering
the expression of the Yukawa correction. This entails that it is possible 
to write the constraints directly in terms of this anomalous acceleration 
$\delta A$. 

This fact is especially clear for the Kepler test since the condition (\ref%
{condEps}) can be rewritten 
\begin{equation}
\varepsilon <\left\vert \frac{\delta A\left( a_{\mathrm{Mars}}\right) }{%
A\left( a_{\mathrm{Mars}}\right) }-\frac{\delta A\left( a_{\mathrm{Earth}%
}\right) }{A\left( a_{\mathrm{Earth}}\right) }\right\vert. 
\end{equation}%
Now $\delta A\left( a\right) $ has essentially the same value at $a=a_{%
\mathrm{Mars}}$ and $a=a_{\mathrm{Earth}}$ as well as at any distance
smaller than the Yukawa range. If we simply denote this constant anomalous
acceleration as $\delta A$, we deduce that a test with a relative accuracy $%
\varepsilon $ is immediately translated into a bound on $\delta A$ 
\begin{equation}
\left\vert \delta A\right\vert <\varepsilon \left\vert \frac{A_{\mathrm{Earth%
}}A_{\mathrm{Mars}}}{A_{\mathrm{Earth}}-A_{\mathrm{Mars}}}\right\vert 
\label{deltaKepler}.
\end{equation}%
Considering\cite{AndersonYukawa} that the distance to Mars has been tested
with an accuracy of the order of 12m, we obtain $\varepsilon \sim 10^{-10}$.
Inserting the values of $A_{\mathrm{Mars}}\sim 2.6\times 10^{-3}\unit{m}%
\unit{s}^{-2}$ and $A_{\mathrm{Earth}}\sim 5.\,9\times 10^{-3}\unit{m}\unit{s%
}^{-2}$, we deduce that $\delta A$ should remain smaller than $5\times
10^{-13}\unit{m}\unit{s}^{-2}$. This is certainly much smaller than the
value (\ref{Pioneer}) needed to explain the Pioneer anomaly. 

For the perihelion test, the accuracy (\ref{precessAccur}) can be rewritten
in the long range limit $r\ll \lambda $ as 
\begin{equation}
\left\vert \frac{\delta \varphi }{2\pi }\right\vert =\left\vert \frac{\delta
A\left( a_{\mathrm{Mars}}\right) }{A\left( a_{\mathrm{Mars}}\right) }%
\right\vert. 
\end{equation}%
It follows that a test with an accuracy $\left\vert \delta \varphi
\right\vert $ is again translated into a bound on $\delta A$ 
\begin{equation}
\left\vert \delta A\right\vert <\left\vert \frac{\delta \varphi }{2\pi }%
\right\vert \left\vert A_{\mathrm{Mars}}\right\vert   \label{deltaPeri}.
\end{equation}%
Both results (\ref{deltaKepler}) and (\ref{deltaPeri}) constrain the value
of $\left\vert \delta A\right\vert $ that is also the combination $\alpha
/\lambda ^{2}$ of the two Yukawa parameters. One can therefore extract the
best bound on $\left\vert \delta A\right\vert $ from the long range
asymptote on the diagram $\left( \lambda ,\alpha \right) $ of Fig.~1. This
diagram collects the most recent planetary data\cite{Coy03} and it leads to
the following bound 
\begin{equation}
\delta A\lesssim 3\times 10^{-13}\unit{m}\unit{s}^{-2}.
\end{equation}

We can now sum up the present paper as discarding the
possibility that the Pioneer anomaly could be explained from a long-range
Yukawa deviation from the Newton potential. This conclusion\cite%
{AndersonYukawa} has been written directly in terms of the anomalous
acceleration $\delta A$ which appears in such a modification of Newton law.
The value needed to fit the Pioneer anomaly is in fact more than 1000 times
too large to remain unnoticed on tests of the Kepler law or of the
precession of perihelions. The discrepancy illustrates the challenge to be
met when trying to analyze the Pioneer anomaly in the same framework as
other tests of gravity.\cite{JaekelTBP}

\section*{Acknowledgments}

We thank H. Dittus, E. Fischbach, R. Hellings, A. Lambrecht, M. N. Nieto, P.
Touboul and S. G. Turyshev for stimulating discussions. We acknowledge the
use of unpublished material kindly transmitted by the authors of Ref.~18.

\end{document}